\documentclass[a4paper,10pt]{article}
\usepackage[utf8]{inputenc}

\usepackage{amsmath}
\usepackage{amssymb}
\usepackage{algorithmic}
\usepackage{fancyvrb}

\newcommand{\cB}{\mathcal{B}}
\newcommand{\cC}{\mathcal{C}}
\newcommand{\cD}{\mathcal{D}}

\newcommand{\djt}{\vee}
\newcommand{\cjt}{\wedge}
\newcommand{\ant}{\ast}
\newcommand{\Mat}{\operatorname{Mat}}

\newcommand{\code}[1]{{\tt #1}}

\title{Fast algorithms for anti-distance matrices 
as a generalization of Boolean matrices}
\author{Michiel de Bondt}

\begin{document}

\maketitle

\begin{abstract}
We show that Boolean matrix multiplication, computed as a sum of products 
of column vectors with row vectors, is essentially the same as Warshall's 
algorithm for computing the transitive closure matrix of a graph from
its adjacency matrix.

Warshall's algorithm can be generalized to Floyd's algorithm for 
computing the distance matrix of a graph with weighted edges. We will
generalize Boolean matrices in the same way, keeping matrix multiplication
essentially equivalent to the Floyd-Warshall algorithm. This way, we get
matrices over a semiring, which are similar to the so-called 
``funny matrices''. 

We discuss our implementation of operations on Boolean matrices and 
on their generalization, which make use of vector instructions.
\end{abstract}

\section{Introduction}

In \cite[pp.\@ 200-206]{MR0413592}, it has been proved that Boolean matrix 
multiplication has the same time complexity as computing the reflexive 
transitive closure matrix which corresponds to a Boolean matrix $A$ as 
adjacency matrix of a graph, i.e.\@ 
\begin{equation} \label{rtc}
I \djt A \djt (A \cjt A) \djt (A \cjt A \cjt A) \djt \cdots 
\end{equation}
where $I$ is the identity matrix.
More generally, it has been proved that matrix multiplication over a semiring
has the same time complexity as computing \eqref{rtc}

The computation of the transitive closure matrix which corresponds $A$, i.e.
\begin{equation} \label{tc}
A \djt (A \cjt A) \djt (A \cjt A \cjt A) \djt \cdots 
\end{equation}
has the same complexity as well. Namely, we can get \eqref{rtc} from 
\eqref{tc} by addition of $I$, and we can obtain \eqref{tc} from 
\eqref{rtc} by multiplication with $A$. More generally, matrix multiplication 
over a semiring has the same time complexity as computing \eqref{tc}.

We connect Boolean matrix multiplication and taking transitive closure in
another way. Namely, we show that Boolean matrix multiplication, computed 
as a sum of products of column vectors with row vectors, is essentially
the same as Warshall's algorithm for computing the transitive closure
matrix.

Warshall's algorithm can be generalized to Floyd's algorithm for 
computing the distance matrix of a graph with weighted edges. We will
generalize Boolean matrices in the same way, keeping matrix multiplication
essentially similar to the Floyd-Warshall algorithm. This way, we get
matrices over a semiring, which we call ``anti-distance matrices''.

Our anti-distance matrices are similar to the distance-matrices in
\cite[pp.\@ 200-206]{MR0413592} and the ``funny matrices'' in 
\cite{MR1462723}. This is not very surprising, because these 
matrices are used to compute the shortest paths for all pairs
of vertices, briefly APSP (All Pairs shortest Path).

Furthermore, we discuss our implementations of matrix operations
on Boolean matrices and anti-distance matrices. This includes the 
Floyd-Warshall algorithm. Alternatives to the Floyd-Warshall algorithm may
involve some sort of graph traversal. This leads to overhead, and is harder
to parallellize. But parallellization is indeed possible, namely with 
GPU shaders, see e.g.\@ \cite{Harish}.

But we will implement our matrix operations in a simpler way, namely by 
way of vector instructions of the CPU. Using such instructions, the 
Floyd-Warshall-algorithm and our matrix multiplication have been implemented 
more or less in \cite{Han:2006:PGA:1152154.1152189}, where our matrix 
multiplication is called Floyd-Warshall-abc, with the three letters referring 
to the three distinct parameters in $C = A \ant B$.

The authors of \cite{Han:2006:PGA:1152154.1152189} wrote a program which
writes the instructions for the matrix operation. We choose a simpler 
approach, namely we wrote g++ template header files with inline assembly
instructions. Furthermore, the authors of \cite{Han:2006:PGA:1152154.1152189}
use signed edge weights without saturation, where we use unsigned edge weights
with saturation.

\section{Boolean matrix multiplication and Warshall's algorithm}

Let $\cB = \{0,1\}$ denote the Boolean set in numerical fashion. 
We define an addition on $\cB$, namely
$$
\djt : \cB \times \cB \rightarrow \cB
\qquad \mbox{by} \qquad a \djt b = \max\{a,b\}
$$
Furthermore, we define a multiplication on $B$, namely
$$
\cjt : \cB \times \cB \rightarrow \cB
\qquad \mbox{by} \qquad a \cjt b = \min\{a,b\}
$$
Using this addition and multiplication, we can define matrix
addition and multiplication over $\cB$, namely in the same way
as usual. So if we take $A, B \in \Mat_{R,C}(\cB)$, then
$$
(A \djt B)_{ij} = A_{ij} \djt B_{ij} \qquad (1 \le i \le R, 1 \le j \le C)
$$
If we take $A \in \Mat_{R,C}(\cB)$ and $B \in \Mat_{C,C'}(\cB)$, then
$$
(A \ant B)_{ij} = (A_{i1} \cjt B_{1j}) \djt \cdots \djt  (A_{iC} \cjt B_{Cj})
\qquad (1 \le i \le R, 1 \le j \le C')
$$
Let $A_{{\updownarrow}k}$ denote the $k$-th column of $A$ and $B_{k{\leftrightarrow}}$
denote the $k$-th row of $B$. Write $0_{R \times C}$ for the zero matrix with $R$ rows
and $C$ columns. We can compute $P = A \ant B$ by way of the following code:
\begin{center}
\begin{algorithmic}
\STATE{$P := 0_{R \times C}$}
\FOR{$k := 1$ \TO $C$}
\STATE{$P := P \vee (A_{{\updownarrow}k} \ant B_{k{\leftrightarrow}})$}
\ENDFOR
\RETURN{$P$}
\end{algorithmic}
\end{center}

We can interpret a square Boolean matrix, say $A \in \Mat_{D,D}(\cB)$, as the adjacency matrix 
of a directed graph. Let $B \in \Mat_{D,D}(\cB)$ be the adjacency matrix for a directed graph
with the same vertices, but different edges as the graph of which $A$ is the adjacency
matrix. Then $A \djt B$ is adjacency matrix of the union of $A$ and $B$. The zero matrix
$0_{D \times D}$ is the neutral element of $\djt$, and the interpretation of $0_{D \times D}$ is the 
empty graph. 

The interpretation of $A \ant B$ is a little more difficult. In the associated graph, 
there is an edge from $i$ to $j$, if and only if there is a vertex $k$ such that there is a 
$B$-edge from $i$ to $k$ and an $A$-edge from $k$ to $j$. The identity matrix $I_D$ of size 
$D$ is the neutral element of $\ant$, and the interpretation of $I_D$ is a graph with $D$
edges, namely an edge of every vertex to itself.

The matrix which corresponds to the transitive closure of the graph is the infinite sum
\begin{equation} \label{trcl}
A \djt (A \ant A) \djt (A \ant A \ant A) \djt \cdots 
\end{equation}
which is actually just 
$$
A \djt (A \ant A) \djt (A \ant A \ant A) \djt \cdots \djt A^{\ant D'}
= A \ant (I_{D'} \djt A)^{\ant (D'-1)}
$$
for every $D' \ge D$, where $A^{\ant D'}$ is the product of $D'$ copies of $A$. 

$(I_{D'} \djt A)^{\ant (D'-1)}$ can be computed by way of repeated squaring, 
but a better method to compute \eqref{trcl} is Warshall's algorithm:
\begin{center}
\begin{algorithmic}
\STATE{$T := A$}
\FOR{$k := 1$ \TO $D$}
\STATE{$T := T \vee (T_{{\updownarrow}k} \ant T_{k{\leftrightarrow}})$}
\ENDFOR
\RETURN{$T$}
\end{algorithmic}
\end{center}
Notice that Warshall's algorithm is very similar to the multiplication
algorithm.

\section{Anti-distance matrices and Floyd's algorithm}

If we compare the Warshall algorithm with the multiplication algorithm, 
we see that for the two input matrices $A$ and $B$ and the output matrix $P$
in the multiplication algorithm, the same matrix $T$ is taken in the Warshall
algorithm. We will generalize the Booleans matrices in such a way, that
Warshall's algorithm becomes Floyd's algorithm for distances.

We do this in a way to preserve the interpretations of $A \djt B$, 
$0_{D \times D}$, $A \ant B$, and $I_D$ as much as possible. For that
reason, a $1$ indicates distance $0$, and a $0$ indicates distance $\infty$.
Any number in between $0$ and $1$ indicates a distance between $0$ and $\infty$,
i.e.\@ a proper distance. 

So let us replace the Boolean set $\cB$ by the 
interval $\cC = [0,1]$. Let $\delta : \cC \rightarrow [0,\infty]$ denote 
the distance associated to an element of $\cC$. Then $\delta(0) = \infty$ and
$\delta(1) = 0$, so it is natural to impose that $\delta$ is decreasing.

The addition operator on $\cC$ should take the minimum distance, which is 
accomplished by taking the maximum value, because $\delta$ is decreasing.
So we can take $\djt$ as the addition operator for $\cC$, just as for $\cB$. 

The multiplication operator on $\cC$ should add distances, but $\cjt$ only
does this for the distances $0$ and $\infty$, which are in fact the only
distances in the Boolean case. So we cannot preserve $\cjt$. No, we use
another multiplication on $\cC$, namely we define
$$
\ant : \cC \times \cC \rightarrow \cC
\qquad \mbox{by} \qquad a \ant b = \delta^{-1}(\delta(a)+\delta(b))
$$
If we extend $\cjt$ to $[0,\infty]$, then
\begin{align*}
(a \djt b) \ant c &= \delta^{-1} \big( \delta(a \djt b) + \delta(c) \big) \\
&= \delta^{-1} \big( \big(\delta(a) \cjt \delta (b)\big) + \delta(c) \big) \\
&= \delta^{-1} \big( \big(\delta(a) + \delta(c) \big) \cjt 
   \big( \delta(b) + \delta(c) \big) \big) \\
&= \delta^{-1} \big( \delta(a) + \delta(c) \big) \djt 
   \delta^{-1} \big( \delta(b) + \delta(c) \big) \\
&= (a \ant c) \djt (b \ant c)   
\end{align*}
and similarly $c \ant (a \djt b) = (c \ant a) \djt (c \ant b)$, so the 
distributive law is fulfilled. From this, the distributive law for matrices follows.

We can take $\delta(a) = -\log(a)$, where $\log$ is the logarithm with respect
to some base. But infinite intervals and functions like $\log$ and $\exp$ are
not the most convenient things in computer practice. If we know in advance that 
distances will not be in the interval $[S,\infty)$ for some positive number $S$, 
or we are just not interested in distinguishing distances in the interval 
$[S,\infty]$, we can take $\delta(a) = S \cdot (1 - a)$. 

We can go even further. Since continuous intervals are not so convenient in computer 
practice, we can replace $\cC = [0,1]$ by $\cD = \{0,\tfrac1S,\tfrac2S,\ldots,1\}$.
Then $\delta(\cD) = \{0,1,2,\ldots,S\}$. We call $S$ is the saturation distance, and
define
$$ 
\ant : \cD \times \cD \rightarrow \cD \qquad \mbox{by} \qquad a \ant b = \max\{a + b - 1, 0\}
$$
which is just as above with $\delta(a) = S \cdot (1 - a)$. 

In \cite[pp.\@ 200-206]{MR0413592} and \cite{MR1462723}, similars things have been 
done. But the deduction from adjacency matrices is missing. Furthermore, operations
are done directly on $\delta(\cC) = [0,\infty]$ instead of $\cC$, with $\cjt$ 
as the addition operator and $+$ as the multiplication operator. That $+$ plays 
the role of multiplication leads to ``funny matrix multiplication'', where the 
associated identity matrix is as funny as
$$
\left( \begin{array}{cccc}
0 & \infty & \cdots & \infty \\
\infty & 0 & \ddots & \vdots \\
\vdots & \ddots & \ddots & \infty \\
\infty & \cdots & \infty & 0
\end{array} \right)
$$
The neutral matrix for addition by way of $\cjt$ is the matrix of which all entries 
are $\infty$ instead of the zero matrix.

\section{Fast algorithms for anti-distance matrices}

In \cite[pp.\@ 200-206]{MR0413592}, it is proved that the computation of 
\eqref{trcl} has the same time complexity as the computation of $A \ant B$,
both for Boolean matrices and anti-distance matrices, where $A$ and $B$ are 
square matrices of the same size. 

Note that sub-cubic multiplication techniques such as Strassen's algorithm cannot be 
applied directly here, because the addition operator ($\djt$ or $\cjt$) has no inverse.
This can be counteracted, leading to algorithms which may have better asymptotic
behavior, but which may not be very fast in practice, especially if the 
matrices are not extemely large.

For arithmetic with Boolean matrices, the first optimization is due to the fact that
$64$ bits fit into a register of a modern CPU. Namely, we can do arithmetic on $64$
matrix entries in parallel. In {\sf matBool.h}, matrix operations are implemented,
where each matrix row is divided into blocks of $64$ bits each. Below is our code
for the matrix multiplication, as $\code{mr} := (\code{*this}) \ant \code{m}$:
\begin{Verbatim}[fontsize=\small]
template <int R, int C>
template <int CC>
matBool<R,CC> matBool<R,C>::operator * (const matBool<C,CC> &m) const
{
    matBool<R,CC> mr = zeroMatrix;
    for (int c=C; --c>=0; ) {
        blockType *m_c = m.block + c * m.Cblocks; // row c of m
        for (int r=R; --r>=0; ) {
            bool e = ((unsigned char *)(block + r * Cblocks))[c>>3] &
                        (1 << (c & 7)); // entry r,c of *this
            if (e) {
                blockType *mr_r = mr.block + r * mr.Cblocks; // row r of mr
                for (int ccb=m.Cblocks; --ccb>=0; ) {
                    mr_r[ccb] |= m_c[ccb]; // block ccb of row c of m
                }    
            }
        }
    }
    return mr;
}
\end{Verbatim}
The code for the transitive closure is similar.

A subsequent optimization for multiplication is to use the ``four Russians''-algorithm. 
See \cite{Arlazaroff:Economical} or \cite[pp.\@ 243-247]{MR0413592}. 
We did not implement this optimization.
The ``four Russians''-algorithm cannot be applied on the transitive closure algorithm.

In our implementation of the anti-distance matrices, we did not use 
$\cD = \{0,\tfrac1S,\tfrac2S,\ldots,1\}$, but we used $S\,\cD = \{0,1,2,\ldots,S\}$. 
This is to be compatible with standard unsigned integer types. We made implementations
for three values of $S$, namely $2^8 - 1$, $2^{16} - 1$, and $2^{32} - 1$, 
corresponding to the standard integer types \code{unsigned char}, \code{unsigned short},
and \code{unsigned int} respectively.

We use template parameter \code{T}, being one of \code{unsigned char}, \code{unsigned short},
and \code{unsigned int}, to select the corresponding value of $S$. In {\sf matAntidist.h},
matrix operations are implemented \emph{without optimization} by way of vector 
instructions. Below is our code for the matrix multiplication, again as 
$\code{mr} := (\code{*this}) \ant \code{m}$:
\begin{Verbatim}[fontsize=\small]
template <int R, int C, class T>
template <int CC>
matAntidist<R,CC,T> matAntidist<R,C,T>::operator * (const matAntidist<C,CC,T> &m) const
{
    matAntidist<R,CC,T> mr = zeroMatrix;
    for (int c=C; --c>=0; ) {
        T *m_c = m.entry + c * CC; // row c of m
        for (int r=R; --r>=0; ) {
            T e = (entry + r * C)[c]; // entry r,c of *this
            if (e) {
                T *mr_r = mr.entry + r * CC; // row r of mr
                e = ~e;
                for (int cc=CC; --cc>=0; ) {
                    T ee = m_c[cc]; // entry c,cc of m
                    if (ee > e) {
                        ee -= e;
                        if (mr_r[cc] < ee) mr_r[cc] = ee;
                    }    
                }    
            }
        }
    }
    return mr;
}
\end{Verbatim}
Notice that $a + b - 1$ in the definition of $\ant$ is scaled to
$a + b - S$, which is computed as $b - (S - a)$ to reduce the number of 
operations. The code for the transitive closure is again similar.

In the optimization with vector instructions, we use Intel's 
SSE instructions. The SSE instructions on Intel compatible hardware 
are the successor of MMX instructions. The instructions act on 
registers of 128 bits. The unsigned variants of integer type SSE 
instructions interpret such a register as a vector of elements of 
$S\,\cD = \{0,1,2,\ldots,S\}$, with as many coordinates as the 
register size permits.
\begin{center}
\renewcommand{\arraystretch}{1.5}
\begin{tabular}{|c|c|c|}
\hline
S & integer type & vector size \\
\hline
$2^{8} - 1$ & \code{unsigned char} & 16 \\
$2^{16} - 1$ & \code{unsigned short} & 8 \\
$2^{32} - 1$ & \code{unsigned int} & 4 \\
\hline
\end{tabular}
\end{center}
In {\sf matAntidist\_sse.h}, matrix operations are implemented 
\emph{with optimization} by way of vector instructions. Below is our code 
for the matrix multiplication, again as 
$\code{mr} := (\code{*this}) \ant \code{m}$:
\begin{Verbatim}[fontsize=\small]
template <int R, int C, class T>
template <int CC>
matAntidist<R,CC,T> matAntidist<R,C,T>::operator * (const matAntidist<C,CC,T> &m) const
{
    matAntidist<R,CC,T> mr = zeroMatrix;
    for (int c=C; --c>=0; ) {
        block_sse<T> *m_c = m.block + c * m.Cblocks; // row c of m
        for (int r=R; --r>=0; ) {
            T e = ((T *) (block + r * Cblocks))[c]; // entry r,c of *this
            if (e) {
                block_sse<T> *mr_r = mr.block + r * mr.Cblocks; // row r of mr
                block_sse<T> b;
                b.clonenot (e);
                for (int ccb=m.Cblocks; --ccb>=0; ) {
                    block_sse<T> bb = m_c[ccb]; // block ccb of row c of m
                    bb.subsat (b);
                    mr_r[ccb].max (bb);
                }
            }
        }
    }
    return mr;
}
\end{Verbatim}
The above code uses three functions with SSE instructions, namely 
\code{clonenot}, \code{subsat}, and \code{max}. 
The code for the transitive closure is again similar.

In \code{clonenot}, the argument is negated logically and copied to all 
coordinates of the vector, which corresponds to the SSE register.
In \code{subsat}, two SSE registers are interpreted as vectors, and
the second vector is subtracted from the first with saturation. 
In \code{max}, again two SSE registers are interpreted as vectors, and
the first vector is replaced by the second vector on spots where 
the second vector is larger.

Below follows the code for \code{clonenot}, \code{subsat}, and \code{max}
with template argument \code{T} = \code{unsigned char}.
\begin{Verbatim}[fontsize=\small]
template <> 
inline void block_sse<unsigned char>::clonenot (unsigned c)
{
    // clone from byte c to doubleword c
    c *= 0x01010101;
    // negate doubleword c logically
    c = ~c;
    // clone from doubleword c to block *this
    asm (
        "movd %1, %0\n\t"
        "pshufd $0, %0, %0"
        : "=x" (val)
        : "r" (c)
    );
}

template <>
inline void block_sse<unsigned char>::subsat (const block_sse<unsigned char> &b)
{    
    asm (
        "psubusb %1, %0"
        : "+x" (val)
        : "xm" (b.val)
    );
}

template <>
inline void block_sse<unsigned char>::max (const block_sse<unsigned char> &b)
{    
    asm (
        "pmaxub %1, %0"
        : "+x" (val)
        : "xm" (b.val)
    );
}
\end{Verbatim}
In \cite{Han:2006:PGA:1152154.1152189}, the authors use distance
matrices instead of anti-distance matrices, so their cloning function
does not need negation. Furthermore, in their code for 
\code{unsigned short} matrix entries, the authors use \code{punpck*}
instructions to duplicate the number of coordinates with the 
clone value, until the whole SSE register is filled, which is after
three such instructions.

For template argument \code{T} = \code{unsigned int}, there is no
instruction for subtraction with saturation. To overcome this,
we use the formula $a := (a \vee b) - b$ for subtraction with 
saturation.
\begin{Verbatim}[fontsize=\small]
template <>
inline void block_sse<unsigned int>::subsat (const block_sse<unsigned int> &b)
{   
    // psubusd does not exist, so one extra instruction is needed
    // using max and regular subtraction 
    asm (
        "pmaxud %1, %0\n\t"
        "psubd %1, %0"
        : "+x" (val)
        : "xm" (b.val)
    );
}
\end{Verbatim}
For addition of \code{unsigned int}egers with saturation, it is recommended to 
use the formula $a := (a \wedge \neg b) + b$, if $\neg b = S - b$ can be 
precomputed. 
But a more funny addition of four \code{unsigned int}egers with saturation 
is given below.
\begin{Verbatim}[fontsize=\small]
template <>
inline void block_sse<unsigned int>::addsat (const block_sse<unsigned int> &b)
{   
    // paddusd does not exist, so extra instructions are needed
    // using regular addition and fix afterwards
    asm (
        "movdqa %1, %%xmm0\n\t"
        "paddd %1, %0\n\t"
        "pcmpgtd %0, %%xmm0\n\t"
        "por %%xmm0, %0"
        : "+x" (val)
        : "xm" (b.val)
    );
}
\end{Verbatim}

Using distance matrices instead of anti-distance matrices
is possible in combination with saturation as well, namely 
by way of $\wedge$ instead of $\vee$, and addition with 
saturation instead of subtraction with saturation. 
Any anti-distance matrix operation and its corresponding 
distance matrix operation satisfy the rules of De Morgan, e.g.\@
$$
\neg (A \vee B) = (\neg A) \wedge (\neg B)
\qquad 
\neg (A \wedge B) = (\neg A) \vee (\neg B)
$$
for anti-distance matrix addition $\vee$ and distance matrix addition 
$\wedge$, and
$$
\neg (A \ant B) = (\neg A) + (\neg B)
\qquad 
\neg (A + B) = (\neg A) \ant (\neg B)
$$
for anti-distance matrix multiplication $\ant$ and distance matrix multiplication $+$.

\section{Conclusion}

Most of our ideas were already discovered earlier, namely in
\cite{Han:2006:PGA:1152154.1152189}, \cite[pp.\@ 200-206]{MR0413592},
and \cite{MR1462723}. The most important new idea is to use 
saturation instead of signed edge values. 

The saturation idea came in a natural way from the Boolean matrices as
adjacency matrices. Due to saturation, we can deal with graphs in which the 
distance between vertices may be infinite, without the need of using floating 
point arithmetic. 
 
Inspired by the Boolean matrices as adjacency matrices, we used anti-distance
matrices instead of distance matrices. Since others used distance matrices,
we added operations for distance matrices as well, which are $+$ (instead of $\ant$) 
for ``funny matrix multiplication'', and \code{dransclose} and \code{dransclosure} 
(instead of \code{transclose} and \code{transclosure}) for computing the distance 
matrix with minimum loop lengths on the diagonal.

With signed matrix entries, saturation is not possible. This is not a problem 
if the matrix entry type is \code{float}, since saturation does not make any 
sense in that case. Furthermore, the correspondence with the Boolean
case cannot be maintained in a natural way, so distance matrices are recommended 
with signed matrix entries.

Our implementation of the matrix multiplication and the Floyd-Warshall algorithm
is not very suitable for very large matrices, since one has to take 
caching into account for such matrices. The authors of 
\cite{Han:2006:PGA:1152154.1152189} do take caching into account. For even larger
matrices, it is faster to make use of GPU shaders, see 
\cite{Katz:2008:ASL:1413957.1413966}.

We used g++ template headers with inline assembly instructions, which can easily
be added to projects which make use of the matrix operations.
In addition to matrix multiplication and the Floyd-Warshall algorithm,
we implemented the following entrywise matrix operations.
\begin{center}
\renewcommand{\arraystretch}{1.5}
\begin{tabular}{|c|c|c|}
\hline
mathematical & c++ & entrywise \\[-5pt]
operator & operator & computation \\
\hline
$\wedge$ & \code{\&} & $\min\{a,b\}$ \\
$\vee$ & \code{|} & $\max\{a,b\}$ \\
$\veebar$ & \code{\^{ }} & $|a-b|$ \\
$\neg$ & \code{\~{ }} & $S-a$ \\
\hline
\end{tabular}
\end{center}

\bibliographystyle{plain}
\bibliography{antidist}

\end{document}